\documentclass[preprint,authoryear,12pt]{elsarticle}

\usepackage{epsfig}
\usepackage{pdflscape}
\usepackage{longtable}
\usepackage{xcolor,colortbl}
\usepackage{color}

\usepackage[a4paper, total={6.5in, 8.5in}]{geometry}

\usepackage{lineno}

\usepackage{amssymb}

\usepackage[ps2pdf,%
breaklinks=true,%
colorlinks=true,%
pdfauthor={First Author et al.},%
pdftitle={Template for manuscripts in Advances in Space Research}%
]{hyperref}

\journal{Advances in Space Research}

\begin{document}

\begin{frontmatter}



\title{HILDCAA* events between 1998 and 2007, and their related interplanetary magnetic field and plasma values}


\author{Alan Prestes}
\address{Vale do Paraiba University - UNIVAP 12244-000 S\~ao Jos\'e dos Campos, SP, Brazil}
\cortext[cor]{Corresponding author}
\ead{prestes@univap.br}

\author{Virginia Klausner\corref{cor}}
\address{Vale do Paraiba University - UNIVAP 12244-000 S\~ao Jos\'e dos Campos, SP, Brazil}
\ead{virginia@univap.br}

\author{Arian Ojeda Gonzalez}
\address{Vale do Paraiba University - UNIVAP 12244-000 S\~ao Jos\'e dos Campos, SP, Brazil}
\ead{arian@univap.br}

\author{Silvio Leite Serra}
\address{Vale do Paraiba University - UNIVAP 12244-000 S\~ao Jos\'e dos Campos, SP, Brazil}
\ead{silvio-serra@bol.com.br}

\begin{abstract}
We investigate the interplanetary conditions during 135 less strict high-intensity, long-duration, continuous AE activity (HILDCAA*) events between the years 1998--2007.
The HILDCAA* events were chosen by following the three ``traditional'' criteria which describe the high-intensity, long-duration, continuous AE activity (HILDCAA).
However, we include a small modification in the criteria that considers: ``the AE values do not drop below 200 nT for more than 2 h at a time''.
This criteria is modified by changing 2 to 4 hours period in which the AE values should not drop below 200 nT.
Once the events are selected, we perform a statistical analysis of the interplanetary parameters during their occurrences.
The distribution of HILDCAA* events along the solar cycle shows a pattern of double peak, 
with a peak around the maximum of the sunspot cycle, and an other in the descending phase.
This kind of distribution is similar to the distribution of low-latitude coronal holes.
For each of the HILDCAA* events, we have found its related Interplanetary Magnetic Field (IMF) and plasma parameter signatures.
The average values of AE, AU, AL, and Dst indices, the density and temperature of the solar wind protons, the solar wind speed,
the Bz component of the IMF, the IMF intensity, dynamic pressure, and plasma beta, among all the 135 HILDCAA* events, found here are:
AE (348.1 $\pm$ 67.1 nT), AU (123.9 $\pm$  34.2 nT), AL (-224.2 $\pm$ 47.3 nT), Dst (-22.1 $\pm$ 9.1 nT), Density (4.3 $\pm$ 1.3 cm-3),
Temperature (170976.3 $\pm$ 58049.9 K), Flow speed (547.1 $\pm$ 83.9 km/s), Bz (-0.70 $\pm$ 0.88 nT), IMF magnitude average (6.6 $\pm$ 1.2 nT), pressure (2.4 $\pm$ 0.6 nPa),
and plasma Beta (0.66 $\pm$ 0.27).

\end{abstract}


\end{frontmatter}

\parindent=0.5 cm

\section{Introduction}

Different solar wind sources cause variations in the interplanetary medium, 
such as variations in density, speed, temperature, and interplanetary magnetic field (IMF). 
Upon reaching Earth magnetosphere, 
these variations generate disturbances in the magnetospheric and ionospheric current systems changing the geomagnetic field values measured on ground.
Solar activity is responsible for many disturbances in the geomagnetic field that can be recurrent or transient.
In general, there are basically 2 solar sources responsible for geomagnetic variation over a solar cycle.
First: Interplanetary Coronal Mass Ejection (ICME) related magnetic storms near solar maximum, and
second: Corotating Interaction Regions (CIR) storms during the solar minimum and its declining phase
\citep{Tsurutani1987PSS,Tsurutani1997AGU,Tsurutani1988JGR,Tsurutani1995JGR:SP,Gosling1993,Gonzalez1994JGR,Suess1998,Richardson2000JGR,Echer2008JGR:SP}.
The solar activity varies over an 11-year period, 
therefore, the interplanetary medium structures and the interplanetary magnetic field also vary systematically according to the solar activity \citep{Parks1991,Kivelson1995}.

During minimum and descending solar activity, the possibility for geomagnetic disturbance is mainly due to the reccurent of coronal holes HSS (High Speed Stream). 
It is common that the coronal holes persist for more than one solar rotation \citep{Sheeley1976,Sheeley1977, Tsurutani1995JGR:SP}.
Therefore, their recurring effects can be seen on Earth for several days or weeks, lasting up to an entire solar rotation about $\sim$ 27 days \citep{Tsurutani1995JGR:SP}. 
High Speed Streams (HSSs) \citep{SmithandWolfe:1976} are characterized by the presence of Alfv\'en waves propagating from the Sun \citep{BelcherandDavis:1971,Shugai2009}. 
Due to their fast velocity, the HSSs flow more radially from the Sun than the slow solar wind from other regions \citep{Guarnieri2006}.

Alfv\'en waves are often detected a few days after the CIRs, and they can last (but not always) for several days.
The magnetic reconnection between the component in the southern direction of the magnetic field of Alfv\'en waves
and the magnetospheric field is the mechanism to energy transfer between the solar wind and the Earth's magnetosphere.
Sometimes, High-Intensity Long-Duration Continuous AE Activity (HILDCAA) can occur \citep{Tsurutani1995JGR:SP,Tsurutani2004JASTP}.

\cite{Tsurutani1987PSS} defined HILDCAA events as intervals of: (1) High-Intensity - AE peak values exceed 1000 nT; (2) Long-Duration - the durations were greater than 2 days; 
(3) Continuous AE Activity - the AE values never dropped below 200 nT for more than 2 h at a time; and (4) HILDCAAs must occur outside of the main phase of magnetic storms. 
These criteria were imposed to illustrate the physical process related to the aforementioned solar energy transfer into the magnetosphere.
However, the same physical process may occur even one or more of the four criteria are not strictly followed \citep{Tsurutani2004JASTP}.

Here, we are interested in analyzing the solar wind speed, density, intensity and orientation of the interplanetary magnetic field, and temperature during HILDCAA* events.
During geomagnetically quiet periods, the solar wind flows at Earth's orbit with speed of about 400 km/s and density of approximately 5 $cm^{-3}$,
and it carries a magnetic field of about 5 nT. 
The solar wind controls the size of the magnetic cavity, through its dynamic pressure, and the energy flow into the magnetosphere by the magnetic reconnection 
of the interplanetary field with the terrestrial field \citep{Dungey1961}.
The bow shock heats and compresses the magnetospheric plasma.
It also compresses the magnetic field, so Alfv\'en waves in the HSS are amplified as they pass through the bow shock. 
Moreover, the southward component direction of the interplanetary magnetic field is amplified as well.
If there are strong southward components within the CIR, a magnetic storm will happen \citep{Dungey1961}. 

Moreover, we verify the effects of the interactions between the solar wind and the magnetosphere using the geomagnetic indices, AE and Dst, 
which are designed to represent the disturbance level of the Earth's magnetosphere. 
Basically, the goal of any index is to provide information in a continuous manner about a more or less complex phenomena which varies with time. 
An index can be used for two purposes: to study the phenomenon itself, or as a reference to study an associated phenomenon \citep{Mayaud1980}.
Here, we will use the AE and Dst indices to verify the associated interplanetary medium conditions.

\section{Datasets and methodology}

In this paper, we use the same methodology developed and evaluated by \cite{Prestes2016} to the identify HILDCAAs* events.
As discussed by \cite{Prestes2016}, HILDCAA* events are defined by the following criteria: 
(1) AE peak values must exceed 1000 nT at least once during each event;
(2) The event should last at least two days, 
(3) the AE values should never drop below 200 nT for periods longer than 4 hours at a time, and 
(4) HILDCAAs* must occur outside of the main phases of magnetic storms.

In order to determine the typical values of the solar wind parameters related to HILDCAA* events, 
we use ACE spacecraft data set with one-minute temporal resolution. 
The interplanetary solar wind data used here are: solar wind speed (Vsw in km/s), plasma density (Nsw in cm$^-3$ ), temperature (Tsw in K),
flow pressure (Psw in nPa), IMF magnitude (Bo in nT),
the Bx (nT), By (nT), Bz (nT) components in the GSM coordinate system, and plasma beta.
We also use the AE, AU, and AL indices with one-minute temporal resolution, and the Dst index with one-hour resolution.

\begin{figure}[ht]
\centering
 \noindent
\includegraphics[width=13cm]{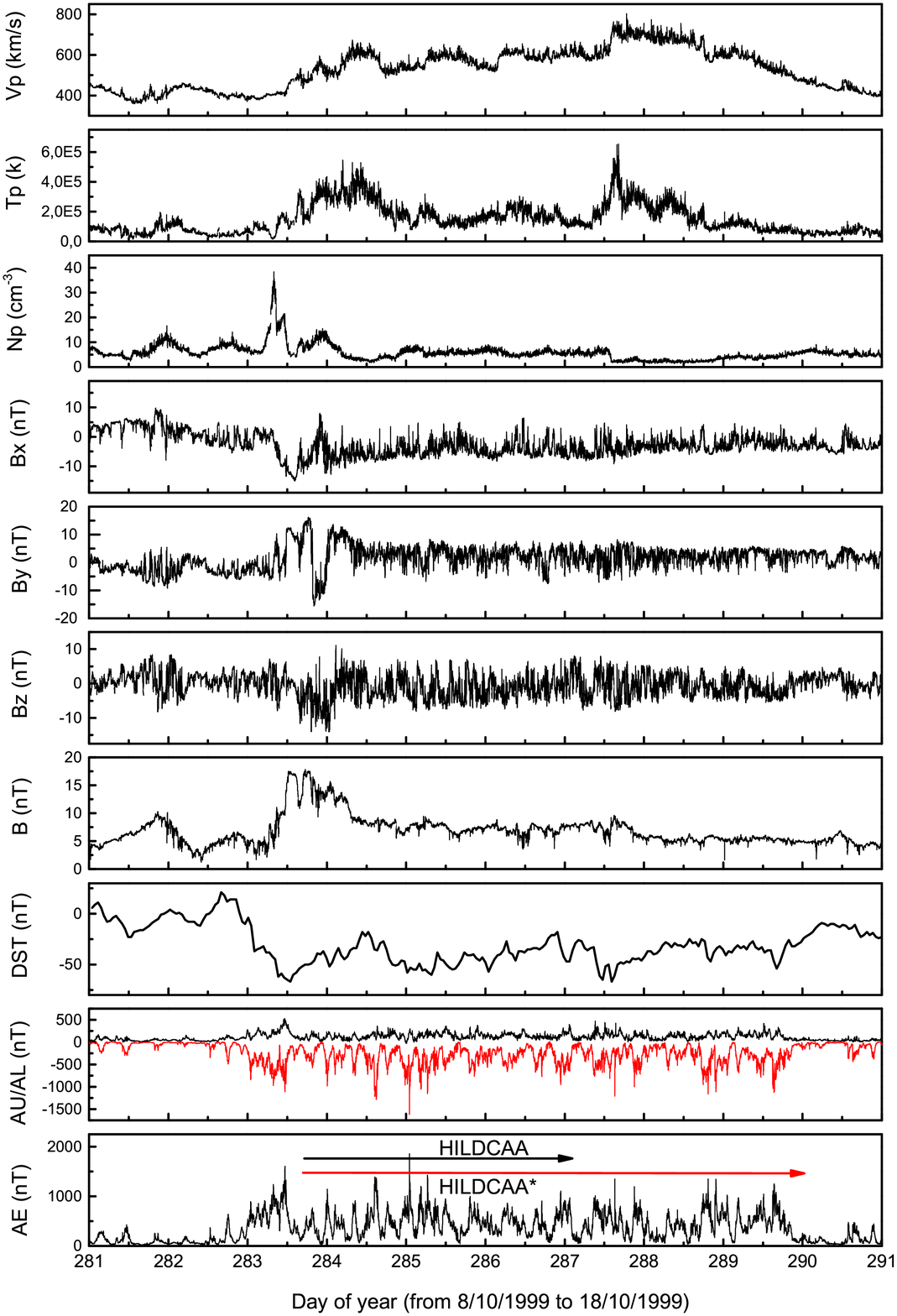}
\caption{From top to bottom: (a) solar wind speed (Vsw in km/s), (b) temperature (K), (c) plasma density (Nsw in cm$^-3$ ), 
(d) Bx component, (e) By component, (f) Bz component in the GSM coordinate system (nT), (g) IMF magnitude (Bo in nT), (h) Dst index (nT), (i) AU/AL indices (nT), and (j) AE index (nT).
Red arrows indicate the interval of the HILDCAA event as defined by \cite{Tsurutani1987PSS}) (beginning at 23:01 of day 10/10 and ending at 17:38 on 14/10),
and HILDCAA* event (beginning 23:01 day 10/10 and ending at 07:57 on 18/10). }
\label{fig:2-1Alan}
\end{figure}

Figure~\ref{fig:2-1Alan} gives an example of ``traditional'' HILDCAA and HILDCAA* events.
It is possible to notice that the HILDCAA event defined by \cite{Tsurutani1987PSS}) is entirely contained within a HILDCAA* event defined by \cite{Prestes2016} (event $10\_1999$).
In this example, both events followed a geomagnetic storm.
For this reason, the onset of HILDCAAs and/or HILDCAAs* is not taken until Dst$>$-50 nT,
and consequently, these events occurred within the storm recovery phase.
As discussed by \cite{Tsurutani1987PSS}, the recovery phase during this kind of events is much longer than usual and can last for more than a week as shown in Figure~\ref{fig:2-1Alan}. 
In addition, it is possible to notice that the AE index shows intense and continuous activity.

\begin{figure}
\noindent
 \includegraphics[height=6.5cm]{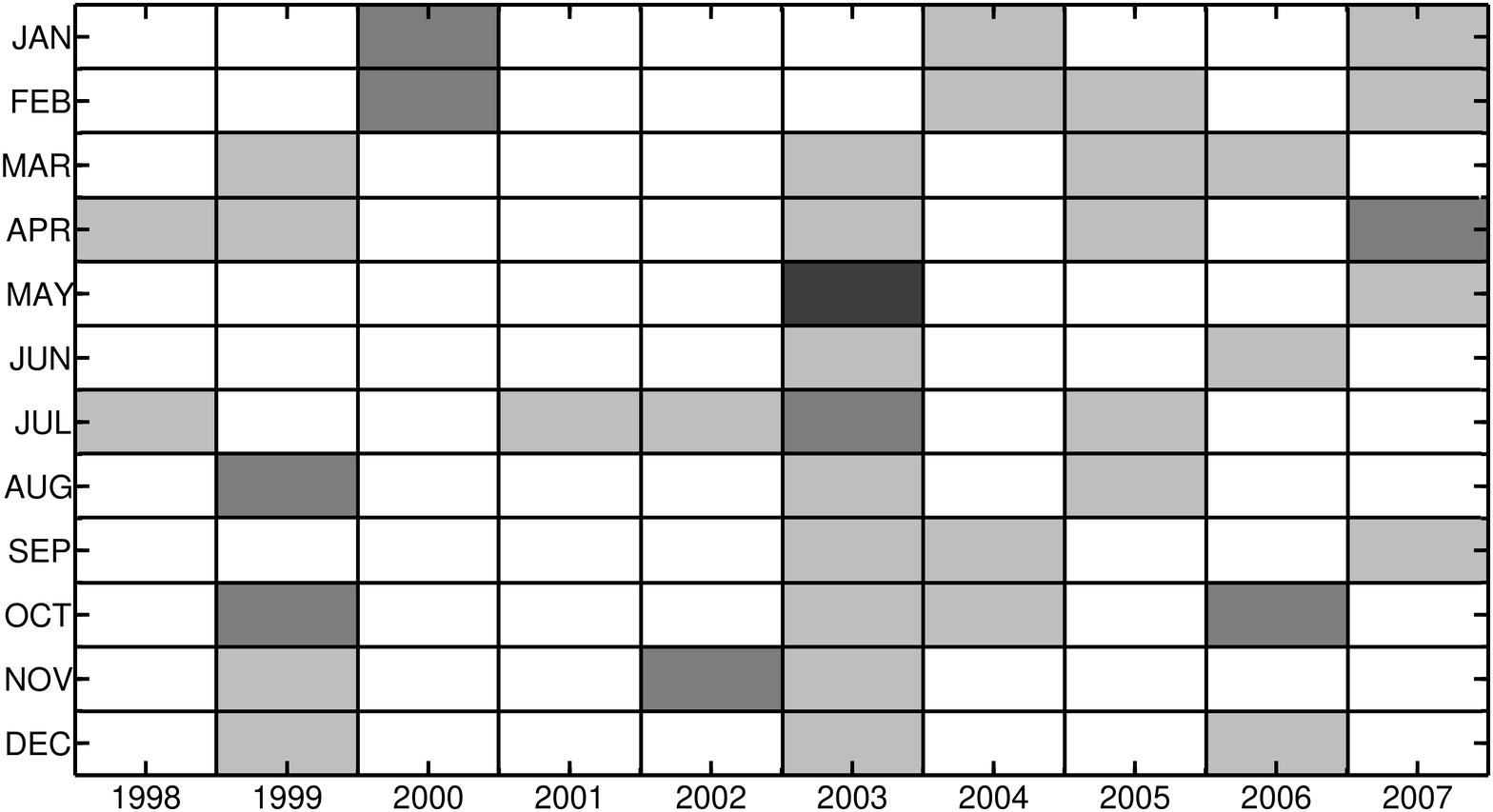}
 \includegraphics[height=6cm]{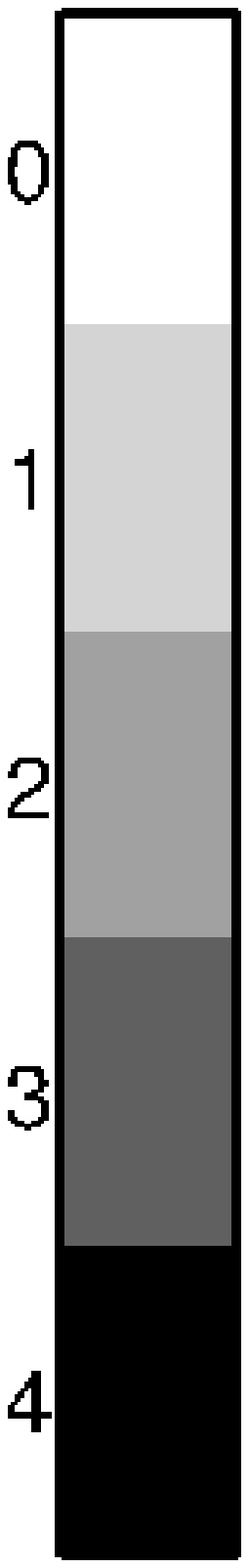}\\
 \includegraphics[height=6.5cm]{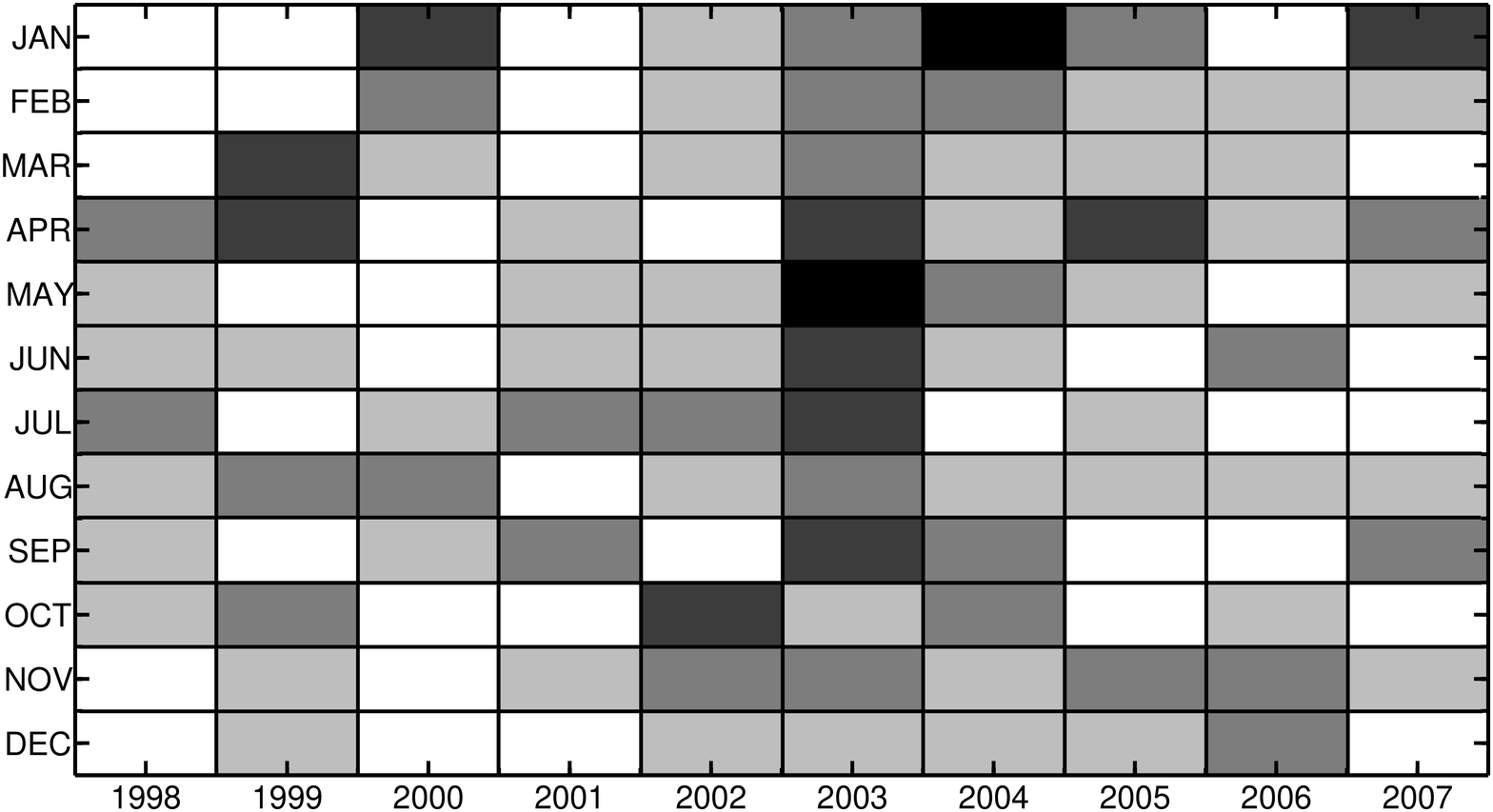}
 \includegraphics[height=6cm]{legenda.eps}\\
 \caption{The number of HILDCAAs (top panel) and HILDCAAs* (bottom panel) for the twelve months of the years 1998--2007.
 The number of events for each month is given by the shades of gray in the legend at the right.}
 \label{Hildcaa}
\end{figure}

Figure~\ref{Hildcaa} shows all of the monthly HILDCAAs (top panel) and HILDCAAs* detected between 1998--2007.
From this figure, it is seen an increase of the event numbers using the criteria defined by \cite{Prestes2016}.
On the top panel, there are only a few regions where there are many events.
On the other hand, on the bottom panel this fact is not true, but the HILDCAA* events present the same variability of HILDCAA events according to the solar activity.
There are a larger occurrence of HILDCAA and HILDCAA* during the declining phase (2003),
a poor occurrence during the high solar activity (2001).

Following the criteria defined by \cite{Prestes2016}, a total of 135 HILDCAA* events were found in the dataset from the years 1998 to 2007.
The list of these events is presented in Table~\ref{grid_mlmmh}, which is arranged as follows: 
column~1 gives the studied events defined by numbers (from 1, 2, 3 and so forth...) and occurrence year;
column~2 gives the starting date;
column~3 gives the starting time in Universal Time (UT);
column~4 gives the ending date; 
column~5 gives the ending time; and 
column~6 gives the preceding and/or associated interplanetary feature.
In column~6, CIR stands for CIR magnetic storm preceding HILDCAA;
CIR? stands for the possibility of CIR magnetic storm preceding HILDCAA;
ICME stands for ICME magnetic storm preceding HILDCAA;
ICME? stands for the possibility of ICME magnetic storm preceding HILDCAA;
ICME+CIR stands for ICME and CIR magnetic storm preceding HILDCAA;
ICME/CIR stands for ICME followed by CIR magnetic storm preceding HILDCAA;
CIR/ICME stands for CIR followed by ICME magnetic storm preceding HILDCAA;
CIR/ICME? stands for the possibility of CIR followed by ICME magnetic storm preceding HILDCAA; and
? indicates that any interplanetary feature was observed.

\section{Results}

\begin{figure}
\centering
\noindent\includegraphics[width=12cm]{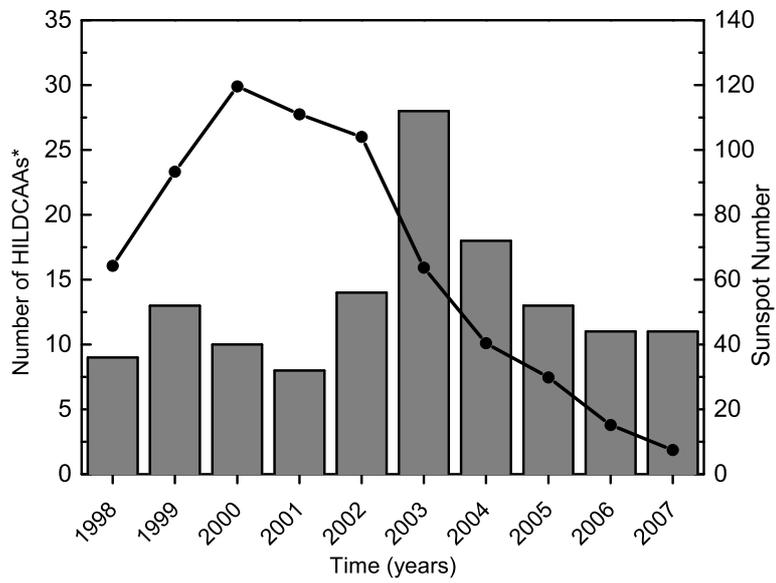}
\caption{Distribution of HILDCAA* events (bars) and annual averages of sunspot number (line) from 1998 to 2007.}
\label{fig:2Alan}
\end{figure}

\begin{figure}
\centering
 \noindent\includegraphics[width=12cm]{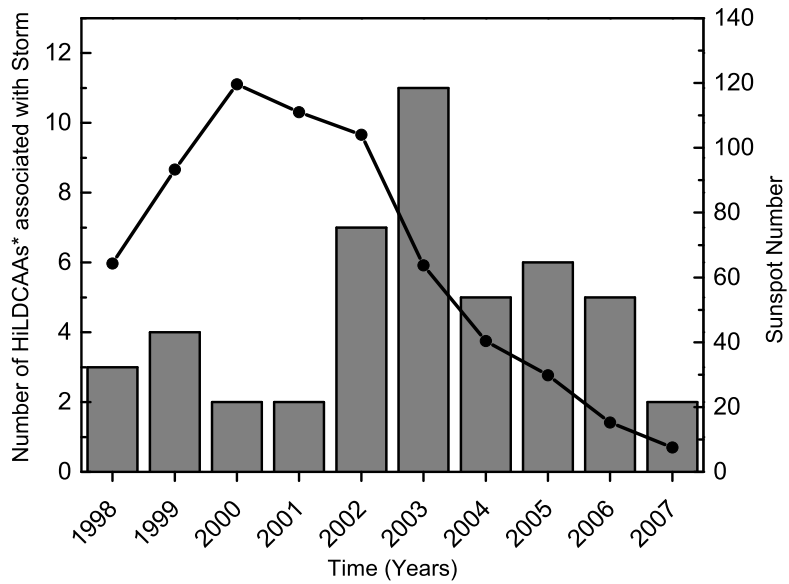}
\caption{Histogram of the HILDCAA* events associated with geomagnetic storms from 1998 to 2007.}
\label{fig:3Alan}
\end{figure}

Figure~\ref{fig:2Alan} shows the HILDCAA* events distribution, and the sunspot number variation over the years 1998--2007. 
Two peaks in the HILDCAA* distribution are observed in this figure.
One: around the maximum of the sunspot cycle,
and another: very intense in the descending phase of the cycle. 
The same kind of feature was observed by \cite{Hajraetal2013JGR}.
From 135 HILDCAA* events occurring between 1998 and 2007, 47 $(\sim35\%)$ followed the storm main phase and occurred during the recovery phase. 
Figure~\ref{fig:3Alan} shows the number of HILDCAA* events which followed geomagnetic storms between the years of 1998 and 2007.

In addition, these 47 storm-associated HILDCAA events are highlighted in gray color in Table~\ref{grid_mlmmh}.
The majority ($\sim$75\% -- CIR (24 events), CIR? (4 events), ICME/CIR (3 events), CIR/ICME (3 events) and CIR+ICME (1 event)) 
of these storm-associated HILDCAA* events are associated with HSS.
Approximately 10\% of the cases (5 events) occurred after the passage of ICMEs.
The remaining 15\% could not be related to any interplanetary feature.
\cite{Hajraetal2013JGR} observed that 94\% of storm-associated HILDCAA events were typically HSS-related, and consequently,
associated with large interplanetary magnetic field (IMF) Bz variances.
As discussed by \cite{Hajraetal2013JGR}, the small percentage of ICME-related HILDCAA events can be explained due to the small,
steady southward Bz intervals or low-frequency fluctuations during the passage of these interplanetary features.

\subsection{Statistical analysis}

For each HILDCAA* event obtained for the period of 1998--2007, the following solar wind parameters are calculated using a statistical approach:
proton density;
proton temperature;
solar wind speed and its components (Vx, Vy, Vz);
magnetic field components (Bx, By, Bz); pressure and plasma beta;
as well as the auroral geomagnetic indices: AE, AU, AL; and the low- to middle-latitude geomagnetic index: Dst. 
Histograms will be present next for all the aforementioned parameters calculated for ten consecutive years (1998--2007).

\subsubsection{AE index}

\begin{figure}
\centering
 \noindent\includegraphics[width=12cm]{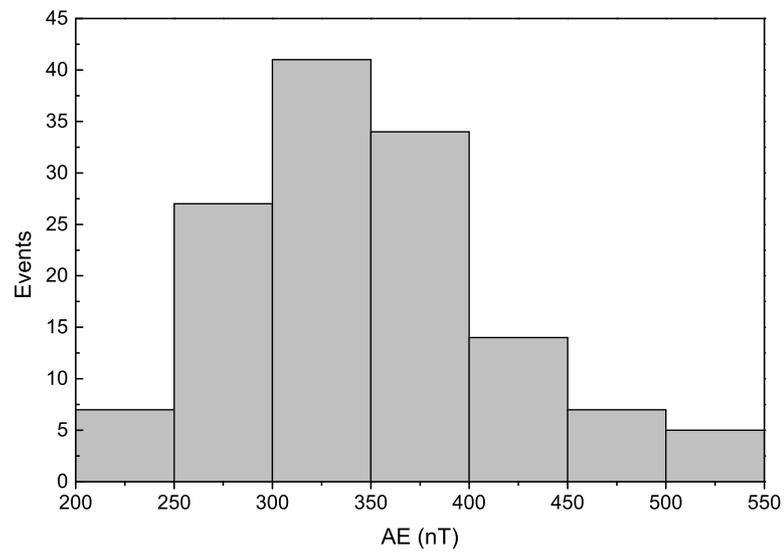}
\caption{Histogram of the mean values of AE index for 135 HILDCAA* events from 1998 to 2007.
Average and standard deviation of the AE index, and Time of occurrence of HILDCAA ($\%$) between 1998-2007. }
\label{fig:4Alan}
\end{figure}

Figure~\ref{fig:4Alan} shows the distribution of AE index mean values for all 135 HILDCAAs* during 1998--2007. 
The average values for these events are above 213.75 nT for the whole period,
and often they reach values above the double of the average value.
The intensity range of the AE average is from a minimum of 224 nT to a maximum of 545 nT.

The majority of the AE average values occurs between 300-400 nT.
In the year of 2003, there are 28 events, almost one every 10 days, and in this year, the AE index has an extremely high average of 328.26 nT, nearly 100 nT higher than  
 the other years of the cycle 23, see Table~\ref{tab:table2}. 
Coronal holes, as well HSSs and CIRs, are a possible source for explain this unsual average value.

\subsubsection{AU/AL indices}

\begin{figure}
\centering
 \noindent\includegraphics[width=12cm]{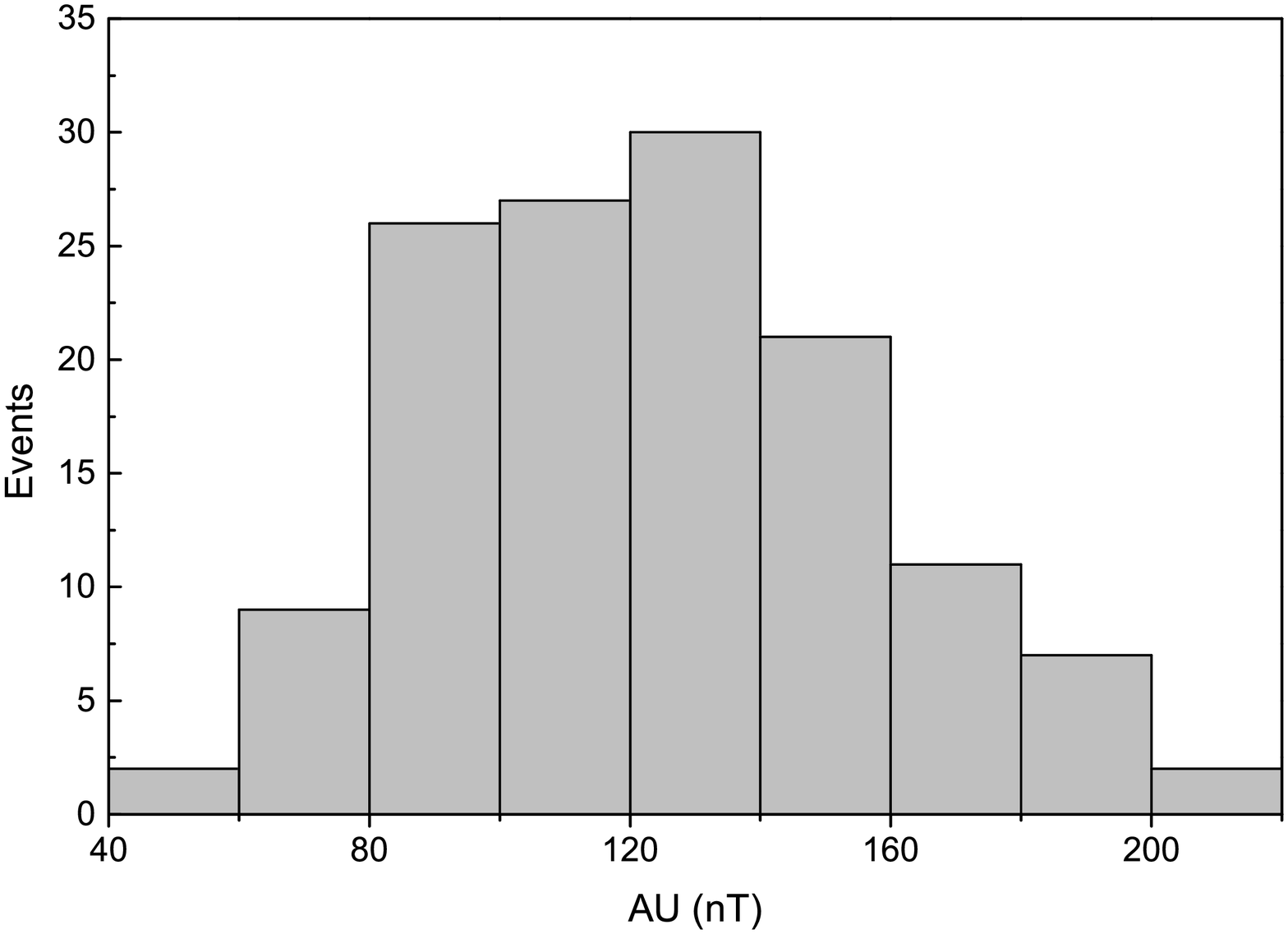}\\
 \noindent\includegraphics[width=12cm]{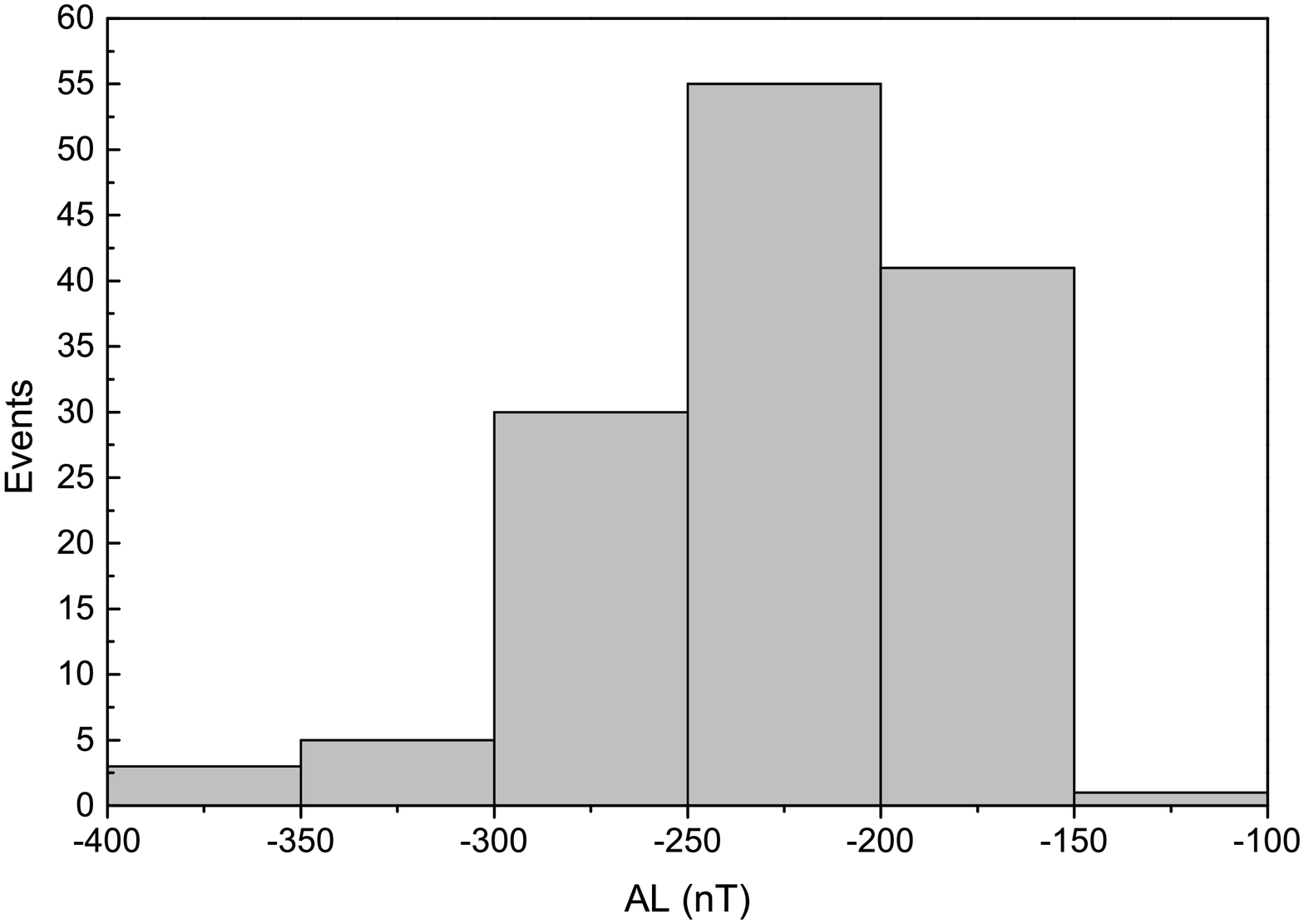}
\caption{Histogram of the average values of the AU (top panel) and AL (bottom panel) indices for all 135 HILDCAA* events between 1998-2007.}
\label{fig:5Alan}
\end{figure}

Figure~\ref{fig:5Alan} shows the histograms of the average values of the AU and AL indices for 135 HILDCAA* events. 
For the AU index, we verify a minimum average of 51 nT and a maximum average of 214 nT.
 
In relation to the AL index during the HILDCAA* events, we verify a minimum average of -393 nT,
and a maximum average of -146 nT. 
The largest number of events presents mean values of AL index between -250 to -150 nT.

\subsubsection{Solar wind proton density}

\begin{figure}
\centering
 \noindent\includegraphics[width=12cm]{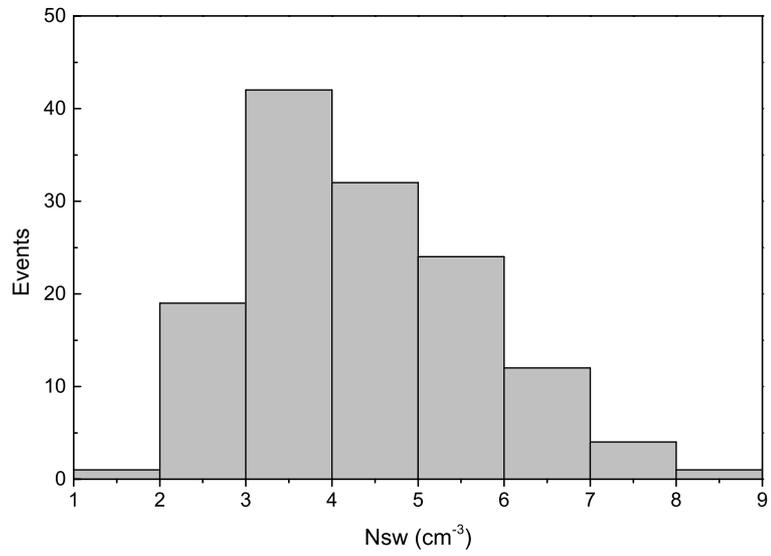}
\caption{Histogram of the average values of the solar wind proton density for all 135 HILDCAA* events from 1998 to 2007.}
\label{fig:6Alan}
\end{figure}

The average values of the solar wind proton density vary 1.9 $cm^{-3}$ to 8.8 $cm^{-3}$,
as noticed in Figure~\ref{fig:6Alan}.
Most of the events shows a density average lower than 6.1 particles per $cm^{3}$.
This kind of results may be related to HSSs.

\subsubsection{Solar wind proton temperature}

\begin{figure}
\centering
 \noindent\includegraphics[width=12cm]{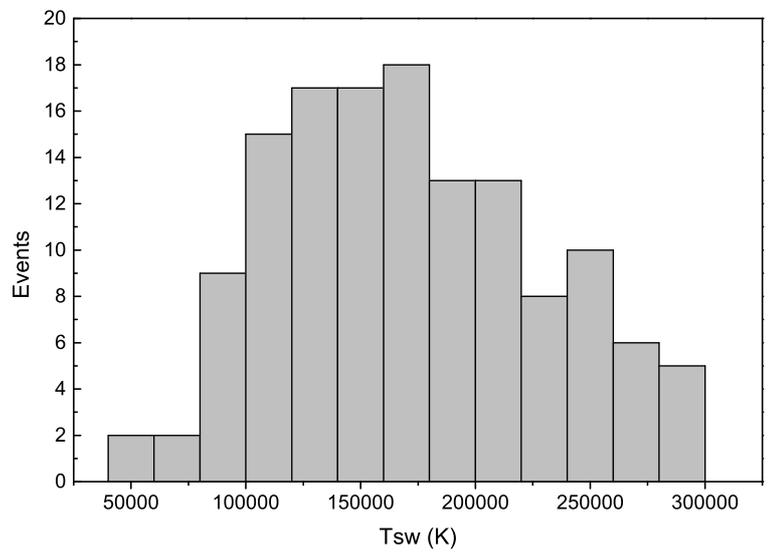}
\caption{Histogram of the average values of the solar wind proton temperature for all 135 events HILDCAA* 1998-2007.}
\label{fig:7Alan}
\end{figure}

The average values of the solar wind proton temperature presented a minimum of 48379 K and a maximum of 298180 K, see Figure~\ref{fig:7Alan}. 
About 85\% of the events  has average values above 125000 K.
Once more, this kind of results may be related to HSSs.

\subsubsection{Solar wind speed}

\begin{figure}
\centering
 \noindent\includegraphics[width=12cm]{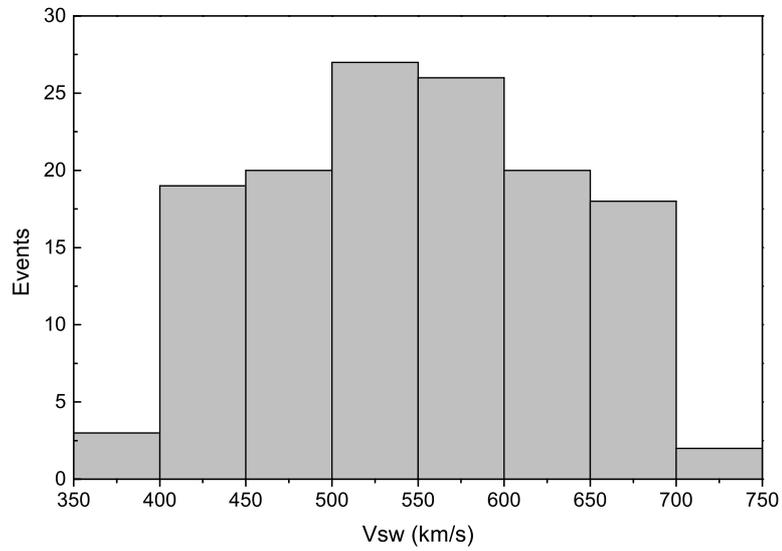}
\caption{Histogram of the average values of the solar wind speed for all 135 HILDCAA* events from 1998 to 2007.}
\label{fig:8Alan}
\end{figure}

Among the 135 events, a minimum average value of the solar wind speed is 369 km/s and a maximum average is 715 km/s, as shown in Figure~\ref{fig:8Alan}.
The $\sim$85\% of the events has average values higher than whole period average which is $\approx$ 450 km/s. 
If we compare to the histogram of the solar wind speed derived from 18 months of ISSE-3 observations \citep[see][page 77, Figure 5]{Russell2001} 
to the solar wind speed histogram in Figure~\ref{fig:8Alan}, 
we verify that the distribution of velocity during the occurrence of HILDCAA* events is shifted 
with respect to the most probable values,
and they are above the level of $75\%$ of the velocity quartiles.

\subsubsection{Bz component of the interplanetary magnetic field}

\begin{figure} 
\centering
 \noindent\includegraphics[width=12cm]{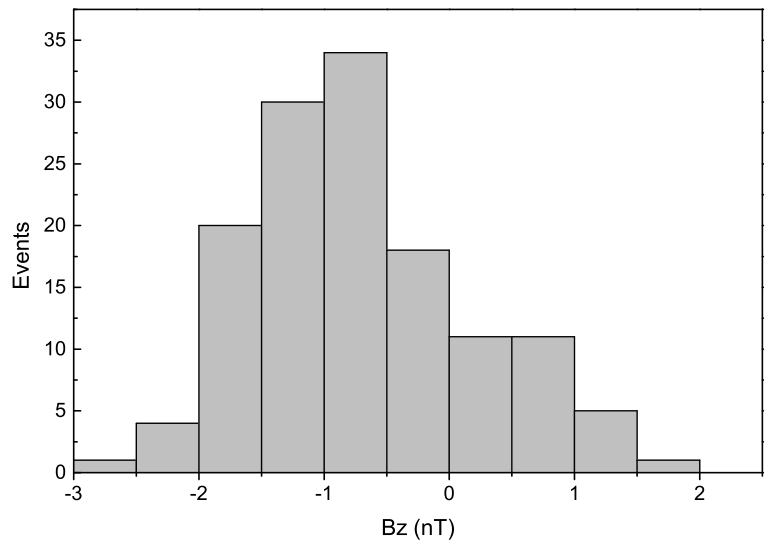}
\caption{Histogram of the average values of Bz (north-south) component of the interplanetary magnetic field for all 135 events HILDCAA* from 1998 to 2007.}
\label{fig:9Alan}
\end{figure}

The Bz component of the interplanetary magnetic field shows mean values between -2.8 nT and 2.0 nT, see Figure~\ref{fig:9Alan}.
The distribution of Bz shows an almost a Gaussian distribution with a slight tendency to the left
different from the distribution of Bz around zero observed by \cite[][page 82, Figure 23]{Russell2001}.
However, when these values is compared to the Bz values during geomagnetic storms, 
we observe that the Bz during HILDCAAs* has smaller values than the average Bz values during storms, 
which is -3 nT for at least 1 hour \citep{Gonzalez1994JGR}.

\subsubsection{Magnitude of solar wind magnetic field}

\begin{figure}
\centering
 \noindent\includegraphics[width=12cm]{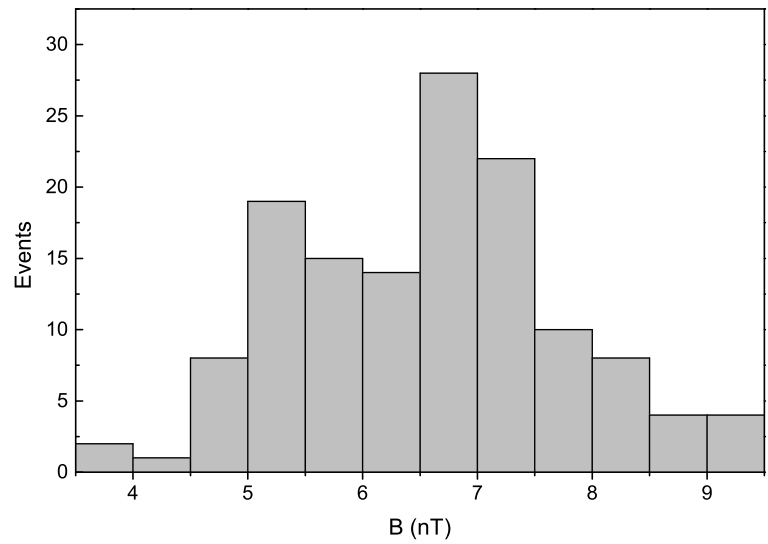}
\caption{Histogram of the average values of the magnetic field magnitude for all 135 HILDCAA* events from 1998 to 2007.}
\label{fig:10Alan}
\end{figure}

The distribution of average values for the interplanetary magnetic field magnitude B is shown in Figure~\ref{fig:10Alan}. 
The occurrence interval of average is between 3.8 nT and 9.4 nT. 
However, the HILDCAA* events occur predominately in the range of 5 to 8 nT.

\subsubsection{Solar wind dynamic pressure}

\begin{figure}
\centering
 \noindent\includegraphics[width=12cm]{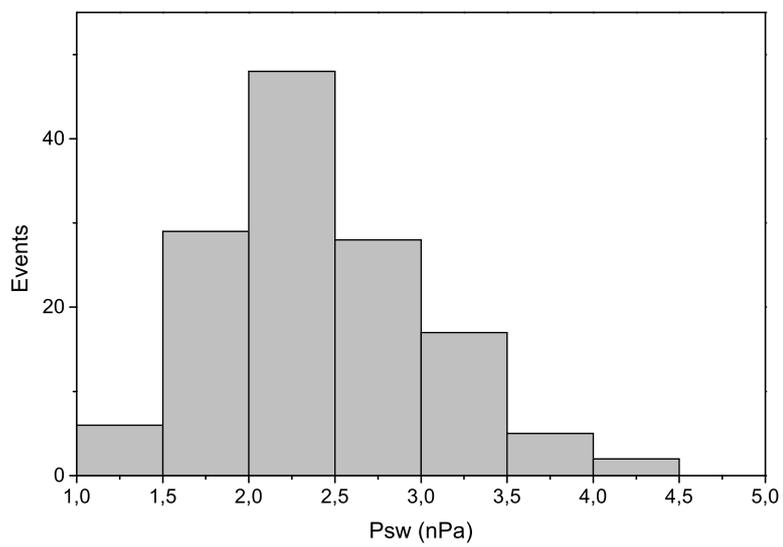}
\caption{Histogram of the average values of the solar wind dynamic pressure for all 135 HILDCAA* events from 1998 to 2007.}
\label{fig:11Alan}
\end{figure}

In the years 1998--2007, during the occurrence of the HILDCAA* events, 
the dynamic pressure has values between 1.2 nPa and 4.5 nPa, as shown in the Figure~\ref{fig:11Alan}.
This histogram illustrates that the interval with the highest frequency of occurrences is between 1.5 to 3 nPa. 
The expected value of the solar wind dynamic pressure at Earth's magnetosphere is about 2.6 nPa at any time of the solar cycle \citep{Kivelson1995}.

The dynamic pressure is similar to high-speed solar wind emerging from the coronal holes,
and of low speed from streamer belt, which does not differ by more than $5\%$. 
Using only this kind of statistical solar wind dynamic pressure analysis, it is not possible to discern any stream structure.
This means that both types of wind have the same dynamic pressure in any obstacle, for example, the planetary magnetospheres \citep{Schwenn2006}.
Therefore, there are no significant differences in the distributions of the dynamic pressure during the occurrence of HILDCAA* events.

\subsubsection{Solar wind plasma beta}

\begin{figure}
\centering
\noindent\includegraphics[width=12cm]{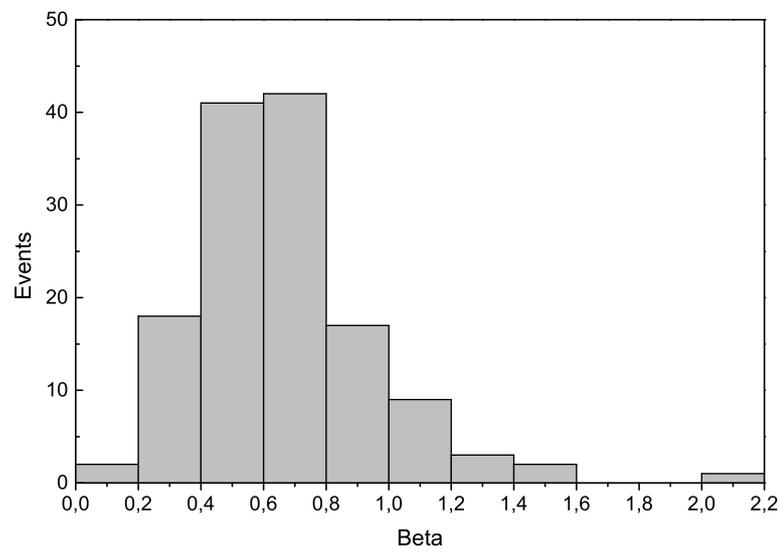}
\caption{Histogram of the average values of solar wind plasma beta for all 135 HILDCAA* events from 1998 to 2007.}
\label{fig:12Alan}
\end{figure}

The $\beta$ parameter is the ratio of thermal pressure to magnetic pressure in the plasma. 
Fast streams have values closer to one, while other structures, such as magnetic clouds or CMEs, have the beta value much smaller than one.
Plasma $\beta$ values are between 0.15 and 2.08, see Figure~\ref{fig:12Alan}. 
However, the $\beta$ values are usually above 0.4 during the HILDCAA* events.

\section{Summary and discussions}

The distribution of HILDCAA* events throughout the solar cycle follows a distribution similar to that observed 
in the distribution of low-latitudes coronal holes during solar cycle 21, as found by \cite[][see Figure 2.6]{Gonzalez1996}.
The existence of two peaks is also observed here.
One: less intense around the solar maximum, and another: more intense in the descending phase.

The time interval of the all analyzed HILDCAA* events varies approximately from 2 days to two weeks.
During the descending phase of the solar cycle, there are an increase of HILDCAA* occurrences,
and also, they persisted for longer time (see Table~\ref{grid_mlmmh}). 
The level of auroral activity, measured by the AE index, has a higher value during this phase, 
possibly due to the presence of long-term fast streams in the solar wind that originate from coronal holes. 
Peak values of AE index between 1000 and 2852 nT are found among all the 135 events, in the period from 1998 to 2007. 
The average values observed are between 224 and 545 nT.

Table~\ref{tab:table3} shows the mean values,with their corresponding standard deviations, for the AE, AU, AL, and Dst indices,
density and temperature of the solar wind protons, solar wind speed, Bz component and intensity of the IMF, the dynamic pressure, and beta factor among all the 135 HILDCAA* events.
\cite{Ahn2003} analyzed the variations of the AU, AL, AE indices with an hourly time series from 1966 to 1987. 
They found average values for AU, AL, AE of  84.6 nT, -135.1 nT, and 213.7 nT (average for all period: both quiet and disturbed), respectively.
For AL index, considering only disturbed period, AL $<$ -135 nT, they found an average of -295.9 nT.
The comparison between the values from Table~\ref{tab:table3} and values found by \cite{Ahn2003} shows that AU during HILDCAA* events presents higher values, 
indicating that the eastward auroral electrojet, from which AU is derived, and it has higher disturbances during the occurrence of HILDCAA* events. 
The mean value of AL for HILDCAAs* is higher than during quiet periods, and lower than during perturbed periods.

\cite{PetrukovichRusanov2005}, who studied the dependence of the AL index in relation to various parameters of the solar wind and IMF,
found that the variables with most influence are in order of importance: (1) electric field (VBs) and speed through viscous interaction; 
(2) By component, solar wind density (AL sensitive to low dynamic pressure), and intensity of IMF fluctuations.
Here, both auroral electrojets, AU/AL, are disturbed during the occurrence of HILDCAA* events. 
This suggests that the two indices are modulated by the same mechanism, possibly by the electric field \citep{Ahn2000,Ahn2008}.

The mean values of AE (348.1 nT), AU (123.9 nT), and AL (-224.2 nT) during HILDCAA* events reached relatively high levels 
if compared with a quiet period (with AE $\approx$ 50 nT),
and with their historical average values of 213.7 nT, 84.6 nT, and -135.1 nT, respectively \citep{Ahn2003}.
The AU and AL indices have distributions similar to AE indicating that during HILDCAA* events both are energized.
But it has been shown by \cite{Rostoker1972,Baumjohann1983,KamideKokubun1996} that the nature of eastward auroral electrojet 
(described by the AU) is quite different from the westward auroral electrojet (described by the AL).
The AE index provides the sum of the maximum current density at two points that are also distant in local time. 
This sum has no physical meaning.

The parameters in the interplanetary medium during HILDCAA* events with highest average values are solar wind speed (547.1 km/s) and temperature (170976.3 K).
Other parameters showed no significant changes, for example B (6.6 nT) and dynamic pressure (2.4 nPa).
The Bz component has a negative average value (-0.70 nT).
However, the expected value is zero. 
Therefore, the preferred direction for this component during HILDCAA* events is southward. 
Remembering that the magnetic reconnection occurs between the southward interplanetary magnetic field component and the geomagnetic field
causing greater injection of energy into the magnetosphere (shown by \cite{Guarnieri2005}).

The plasma beta is useful to identify solar wind structures such as magnetic clouds, shocks, and high-speed solar-wind streams,
and it presents a relatively high average value of 0.66.
Comparing this value to the values related to different structures in the interplanetary medium, 
the plasma beta is associated to high-speed solar-wind streams when its values are close to or larger than one.
Other structures, such as magnetic clouds have a plasma beta values smaller than one. 
This analysis indicates that most events are associated to high-speed streams flowing from coronal holes.
In many cases, these high speed streams are embedded into Alfv\'enic fluctuations covering a wide frequency bandwidth. 
The solar wind Alfv\'enicity has already been indicated by several authors as a possible cause of particle penetration, 
and particle precipitation \citep{Tsurutani1987PSS,Tsurutani1995JGR:SP,Tsurutani1997AGU,Guarnieri2005}.

The distributions of the parameters related to the HILDCAA* events reveals
that only 15 ($\sim 11\%$) of 135 events had average speeds higher than 650 km/s (see Figure~ \ref{fig:8Alan}). 
The histogram in Figure~\ref{fig:6Alan} shows that $75\%$ of events had an average density smaller than the average ($\approx 6.1 cm^{-3}$) from 1998 to 2007. 
Therefore, a large compression from the solar wind to the magnetosphere is not observed 
for most of the events as seen by the low values of the dynamic pressure calculated for all events.
However, it did not exceed 6 nPA, as shown in Figure~\ref{fig:11Alan}. 
For most HILDCAA* events, the average Bz component of the interplanetary magnetic field has mostly negative values (Bz is south)
and a high average speed.
Thus, the electric field Ey given by VBs has a component resulting in the dawn to dusk direction. 
This field configuration describes the energy transfer mechanism of the solar wind into the magnetosphere by magnetic reconnection, 
which is the most efficient mechanism to transfer energy \citep{Gonzalez1994JGR}.

For a better understanding of results for the HILDCAA* events, we compare them to other periods in different conditions to have a reference
between solar wind variations, and their seeding structures in the interplanetary medium.

As discussed by \cite{Schwenn2006}, there are two basic types of solar wind: slow wind and fast wind.
These two kinds of solar wind differ markedly in their main properties,
\textit{i.e.}, in the location  and the magnetic topology of their sources in the solar corona, and probably in the acceleration mechanism.
Some characteristic values of the slow wind are: Vsw between 250--400 km/s, Nsw $\sim$10.7 $cm^{-3}$, Tsw $\sim 3.4\times 10^{4} K$, 
and their sources are found in the streamer belt in the solar corona. 
The characteristic values of the fast solar wind are: Vsw between 400--800 km/s, Nsw $\sim$3.0 $cm^{-3}$, Tsw $\sim 2.3\times 10^{5} K$,
and their sources are in the coronal holes. 

According to \cite{Richardson2000JGR}, the Earth passes about $10\%$ of its time in solar wind structures related to CMEs at solar minimum phase, 
and $60\%$ in solar wind structures related to corotating streams. 
At the solar maximum phase, the Earth passes about $30\%$ of its time in solar wind structures related to CMEs, and $30\%$ in structures related to corotating streams. 
Thus, the Earth spends most of its time in solar wind structures related to fast streams.

\cite{CaneRichardson2003} studied 214 interplanetary coronal mass ejections (ICME), which occurred between 1996-2002,
and they found average values of 454 $\pm$ 6 km/s for velocity and of 9.9 $\pm$ 0.3 nT for Bo. 
The mean speed of these events is less than the average of the HILDCAA* events.
In addition, the average value of the interplanetary magnetic field is higher 
than the values related to HILDCAA* events.
Thus, it can be inferred that HILDCAA* events are not associated with ICMEs.

Moreover, \cite{Gonzalez2011} found average values for Dst peak and some interplanetary parameters associated with superstorms. 
These values were Dst (-324.3$\pm$67.3 nT), Vsw (799.1 $\pm$ 160.6 km/s), Bo (41.7 $\pm$ 10.8 nT), Bz (34.3 $\pm$ 13.5 nT),
Ey (23.5 $\pm$ 11.6 mV/m), Nsw (24.7 $\pm$ 13.7 $cm^{-3}$), and Psw (25.7 $\pm$ 14.8 nPa). 
As can be seen, all parameters reached high values during superstorms.

The results presented and compared above elucidate that some parameters of magnetic field and of plasma of 
the interplanetary medium related to HILDCAA* events are intermediate values between periods of low geomagnetic activity (background values) 
and high geomagnetic activity values (storm values), as summarized in Table~\ref{tab:table4}.
Therefore with the parameters of plasma and IMF on hands, it is possible to determine which kind of geomagnetic event will occur.

\section{acknowledgments}

This research was supported by CNPq and FAPESP: A. Prestes FAPESP - (2009/02907-8) and CNPq (301441/2013-8).
V.~Klausner wishes to thank CNPq for the financial support for her  postdoctoral research  (grants 165873/2015-9).
Furthermore, the authors would like to thank the WDC-Kyoto and OMNIWeb for the data sets used in this work.
The AE and Dst indexes were obtained from the World Data Center for Geomagnetism - Kyoto (http://wdc.kugi.kyoto-u.ac.jp/index.html).
The ACE parameters are available through the National Space Science Data Center
(OMNIWeb):{http://omniweb.gsfc.nasa.gov/form/dx1.html}.



\newpage

\begin{longtable}{l l l l l l l l}
\caption[]{HILDCAA* events for the period between 1998-2007.} \\
\label{grid_mlmmh} \\
\hline \multicolumn{1}{c}{\textbf{Event$\_$Year}}&\multicolumn{1}{c}{\textbf{Start}} 
& \multicolumn{1}{c}{\textbf{UT}}& \multicolumn{1}{c}{\textbf{Stop}}
&\multicolumn{1}{c}{\textbf{UT}}& \multicolumn{1}{c}{\textbf{Length (min)}}& \multicolumn{1}{c}{\textbf{Mechanism}}\\ \hline 
\endfirsthead
\multicolumn{7}{c}%
{{\bfseries \tablename\ \thetable{} -- continued from previous page}} \\
\hline \multicolumn{1}{c}{\textbf{Event$\_$Year}} &
\multicolumn{1}{c}{\textbf{Start}} &
\multicolumn{1}{c}{\textbf{UT}}
&
\multicolumn{1}{c}{\textbf{Stop}}
&
\multicolumn{1}{c}{\textbf{UT}}
&
\multicolumn{1}{c}{\textbf{Length (min)}}
&
\multicolumn{1}{c}{\textbf{Mechanism}}
\\ \hline 
\endhead
\hline 
\multicolumn{7}{r}{{Continued on next page}} \\ \hline
\endfoot
\hline \hline
\endlastfoot
\cellcolor{gray!25}$01\_1998$ &\cellcolor{gray!25}24 Apr &\cellcolor{gray!25}12:00 &  \cellcolor{gray!25}26 Apr &\cellcolor{gray!25}13:00 & \cellcolor{gray!25}2940 & \cellcolor{gray!25}CIR \\
 \cellcolor{gray!25}$02\_1998$ & \cellcolor{gray!25}26 Apr & \cellcolor{gray!25}21:00 & \cellcolor{gray!25}29 Apr  & \cellcolor{gray!25}13:44 &\cellcolor{gray!25}3884 & \cellcolor{gray!25}?\\
 $03\_1998$ & 10 May & 05:32 & 13 May  & 02:24 &4132 &?\\
 $04\_1998$ & 20 Jun & 03:16 & 23 Jun  & 08:50 &4654 &CIR\\
 $05\_1998$ & 03 Jul & 20:00 & 06 Jul  & 17:20 &4160 &CIR\\
 $06\_1998$ & 22 Jul & 20:56 & 25 Jul  & 17:36 &4120 &CIR\\
 \cellcolor{gray!25}$07\_1998$ & \cellcolor{gray!25}28 Aug & \cellcolor{gray!25}22:45 & \cellcolor{gray!25}01 Sep  & \cellcolor{gray!25}05:54 & \cellcolor{gray!25}4749 &\cellcolor{gray!25}ICME\\
 $08\_1998$ & 22 Sep & 19:23 & 24 Sep  & 23:59 &3156 &ICME\\
 $09\_1998$ & 22 Oct & 01:13 & 25 Oct  & 19:01 &5388 &ICME\\
\hline
 $01\_1999$ & 02 Mar & 19:27 & 06 Mar  & 02:14 &4727 &CIR \\
 $02\_1999$ & 13 Mar & 08:34 & 15 Mar  & 23:38 &3785&? \\
 \cellcolor{gray!25}$03\_1999$ & \cellcolor{gray!25}29 Mar & \cellcolor{gray!25}15:01 & \cellcolor{gray!25}01 Apr  & \cellcolor{gray!25}13:34 & \cellcolor{gray!25}4233 &\cellcolor{gray!25}CIR  \\
 $04\_1999$ & 03 Apr & 10:17 & 06 Apr  & 01:50 &3813 &? \\
  \cellcolor{gray!25}$05\_1999$ &  \cellcolor{gray!25}19 Apr &  \cellcolor{gray!25}02:04 & \cellcolor{gray!25} 21 Apr  &  \cellcolor{gray!25}07:16 &  \cellcolor{gray!25}3192& \cellcolor{gray!25}?\\
 $06\_1999$ & 27 Apr & 07:21 & 04 May  & 03:37 &9856 &CIR  \\
 $07\_1999$ & 30 Jun & 19:54 & 03 Jul  & 23:23 & 4529&ICME+CIR \\
 $08\_1999$ & 15 Aug & 22:37 & 20 Aug  & 01:59 &5962 &CIR  \\
 $09\_1999$ & 29 Aug & 08:41 & 03 Sep  & 08:17 &7176  &? \\
 \cellcolor{gray!25}$10\_1999$ & \cellcolor{gray!25}10 Oct & \cellcolor{gray!25}23:01 & \cellcolor{gray!25}18 Oct  & \cellcolor{gray!25}07:57 & \cellcolor{gray!25}10615 &\cellcolor{gray!25}CIR \\
 \cellcolor{gray!25}$11\_1999$ & \cellcolor{gray!25}26 Oct & \cellcolor{gray!25}04:29 & \cellcolor{gray!25}28 Oct  & \cellcolor{gray!25}09:55 & \cellcolor{gray!25}3206 &\cellcolor{gray!25}CIR \\
  $12\_1999$ & 23 Nov & 00:54 & 25 Nov  & 02:49 & 2995 &CIR+ICME\\
 $13\_1999$ & 03 Dec & 01:57 & 10 Dec  & 07:25 &10408  &CIR \\
 \hline
 $01\_2000$ & 01 Jan & 04:00 & 03 Jan  & 20:46 &3886  &CIR?\\
 $02\_2000$ & 04 Jan & 05:46 & 07 Jan  & 21:18 & 5252 &?\\
 \cellcolor{gray!25}$03\_2000$ & \cellcolor{gray!25}27 Jan & \cellcolor{gray!25}18:06 & \cellcolor{gray!25}31 Jan  & \cellcolor{gray!25}23:01 & \cellcolor{gray!25}6056 &\cellcolor{gray!25}CIR\\
 \cellcolor{gray!25}$04\_2000$ & \cellcolor{gray!25}05 Feb & \cellcolor{gray!25}15:53 & \cellcolor{gray!25}08 Feb  & \cellcolor{gray!25}22:16 & \cellcolor{gray!25}4704 &\cellcolor{gray!25}CIR \\
 $05\_2000$ & 24 Feb & 02:42 & 27 Feb  & 21:37 & 5455  &ICME/CIR \\
 $06\_2000$ & 06 Mar & 05:07 & 09 Mar  & 01:03 & 4076 &ICME/CIR \\
 $07\_2000$ & 30 Jul & 11:11 & 02 Aug  & 12:42 &  4411 &CIR \\
 $08\_2000$ & 04 Aug & 00:16 & 07 Aug  & 13:24 &  5108 &?\\
 $09\_2000$ & 30 Aug & 20:39 & 01 Sep  & 22:59 &  3021 &?\\
 $10\_2000$ & 06 Sep & 15:02 & 09 Sep  & 00:18 & 3436  &ICME\\
 \hline
 $01\_2001$ & 06 Apr & 06:37 & 11 Apr  & 12:30 & 7553  &ICME\\
 \cellcolor{gray!25}$02\_2001$ & \cellcolor{gray!25}10 May & \cellcolor{gray!25}11:00 & \cellcolor{gray!25}14 May  & \cellcolor{gray!25}08:47 & \cellcolor{gray!25}5627  &\cellcolor{gray!25}CIR\\
 $03\_2001$ & 09 Jun & 04:00 & 11 Jun  & 05:49 & 2989  &CIR \\
 $04\_2001$ & 15 Jul & 06:19 & 18 Jul  & 08:57 & 4478  &CIR \\
 $05\_2001$ & 21 Jul & 23:40 & 26 Jul  & 13:56 & 6617  &CIR \\
 $06\_2001$ & 02 Sep & 11:29 & 05 Sep  & 03:32 & 3843  &CIR \\
 \cellcolor{gray!25}$07\_2001$ &  \cellcolor{gray!25}27 Sep &  \cellcolor{gray!25}10:17 &  \cellcolor{gray!25}30 Sep  &  \cellcolor{gray!25}15:57 &  \cellcolor{gray!25}4660  & \cellcolor{gray!25}?\\
 $08\_2001$ & 17 Nov & 00:22 & 20 Nov  & 08:10 & 4788  & ?\\
 \hline
 \cellcolor{gray!25}$01\_2002$ & \cellcolor{gray!25}11 Jan & \cellcolor{gray!25}10:00 & \cellcolor{gray!25}14 Jan  & \cellcolor{gray!25}23:11 & \cellcolor{gray!25}5111  &\cellcolor{gray!25}CIR? \\
 \cellcolor{gray!25}$02\_2002$ & \cellcolor{gray!25}06 Feb & \cellcolor{gray!25}11:00 & \cellcolor{gray!25}09 Feb  & \cellcolor{gray!25}13:09 & \cellcolor{gray!25}4449  &\cellcolor{gray!25}CIR\\
 $03\_2002$ & 05 Mar & 00:41 & 07 Mar  & 01:05 & 2904  &CIR/ICME \\
 \cellcolor{gray!25}$04\_2002$ & \cellcolor{gray!25}14 May & \cellcolor{gray!25}13:00 & \cellcolor{gray!25}17 May  & \cellcolor{gray!25}00:01 & \cellcolor{gray!25}3542 &\cellcolor{gray!25}?\\
 $05\_2002$ & 08 Jun & 12:06 & 12 Jun  & 05:25 & 5359  &CIR \\
 $06\_2002$ & 22 Jul & 01:35 & 24 Jul  & 11:58 & 3503  &CIR+ICME\\
 $07\_2002$ & 24 Jul & 18:49 & 28 Jul  & 06:12 & 5004  &CIR+ICME\\
 $08\_2002$ & 13 Aug & 07:00 & 16 Aug  & 17:19 & 4939  &?\\
 \cellcolor{gray!25}$09\_2002$ & \cellcolor{gray!25}09 Oct & \cellcolor{gray!25}01:01 & \cellcolor{gray!25}13 Oct  & \cellcolor{gray!25}13:08 & \cellcolor{gray!25}6487 & \cellcolor{gray!25}ICME/CIR \\
 \cellcolor{gray!25}$10\_2002$ & \cellcolor{gray!25}25 Oct & \cellcolor{gray!25}21:00 & \cellcolor{gray!25}29 Oct  & \cellcolor{gray!25}15:33 & \cellcolor{gray!25}3992  &\cellcolor{gray!25}CIR \\
 $11\_2002$ & 29 Oct & 20:30 & 01 Nov  & 01:48 & 3198  & ?\\
 \cellcolor{gray!25}$12\_2002$ &  \cellcolor{gray!25}04 Nov &  \cellcolor{gray!25}19:00 &  \cellcolor{gray!25}07 Nov  &  \cellcolor{gray!25}20:52 &  \cellcolor{gray!25}4432  & \cellcolor{gray!25}CIR \\
 \cellcolor{gray!25}$13\_2002$ & \cellcolor{gray!25}27 Nov & \cellcolor{gray!25}13:00 & \cellcolor{gray!25}03 Dec  & \cellcolor{gray!25}05:20 & \cellcolor{gray!25}8180  &\cellcolor{gray!25}CIR \\
 $14\_2002$ & 26 Dec & 08:30 & 28 Dec  & 18:25 & 3475  &CIR \\
 \hline
 $01\_2003$ & 19 Jan & 05:46 & 23 Jan  & 05:38 & 5752  &CIR \\
 $02\_2003$ & 23 Jan & 09:44 & 27 Jan  & 02:19 & 5315  &?\\
 \cellcolor{gray!25}$03\_2003$ & \cellcolor{gray!25}06 Feb & \cellcolor{gray!25}05:22 & \cellcolor{gray!25}11 Feb  & \cellcolor{gray!25}18:12 & \cellcolor{gray!25}7970  &\cellcolor{gray!25}CIR \\
 $04\_2003$ & 16 Feb & 05:24 & 21 Feb  & 11:10 & 7546  &CIR/ICME \\
 $05\_2003$ & 04 Mar & 10:00 & 07 Mar  & 23:34 & 5134  &CIR \\
 $06\_2003$ & 14 Mar & 03:01 & 19 Mar  & 21:34 & 8313  &CIR \\
 $07\_2003$ & 03 Apr & 14:11 & 06 Apr  & 10:22 & 4091 &CIR?\\
 $08\_2003$ & 15 Apr & 19:10 & 17 Apr  & 21:06 & 2996  &CIR \\
 $09\_2003$ & 20 Apr & 15:45 & 28 Apr  & 14:31 & 11447  &?\\
 $10\_2003$ & 01 May & 09:00 & 04 May  & 14:35 & 4655  &?\\
 \cellcolor{gray!25}$11\_2003$ & \cellcolor{gray!25}05 May & \cellcolor{gray!25}08:50 & \cellcolor{gray!25}09 May  & \cellcolor{gray!25}08:00 & \cellcolor{gray!25}5710  &\cellcolor{gray!25}CIR/ICME \\
 $12\_2003$ & 10 May & 20:00 & 16 May  & 14:23 & 8303  &ICME\\
 \cellcolor{gray!25}$13\_2003$ & \cellcolor{gray!25}22 May & \cellcolor{gray!25}09:00 & \cellcolor{gray!25}29 May  & \cellcolor{gray!25}11:27 & \cellcolor{gray!25}10227  &\cellcolor{gray!25}CIR? \\
 $14\_2003$ & 02 Jun & 17:00 & 12 Jun  & 18:25 & 14485 &ICME/CIR\\
 \cellcolor{gray!25}$15\_2003$ & \cellcolor{gray!25}13 Jun & \cellcolor{gray!25}18:58 & \cellcolor{gray!25}16 Jun  & \cellcolor{gray!25}07:00 & \cellcolor{gray!25}3602  &\cellcolor{gray!25}ICME/CIR\\
 \cellcolor{gray!25}$16\_2003$ & \cellcolor{gray!25}20 Jun & \cellcolor{gray!25}04:53 & \cellcolor{gray!25}06 Jul  & \cellcolor{gray!25}17:28 & \cellcolor{gray!25}23795  &\cellcolor{gray!25}ICME/CIR\\
 \cellcolor{gray!25}$17\_2003$ & \cellcolor{gray!25}12 Jul & \cellcolor{gray!25}20:00 & \cellcolor{gray!25}15 Jul  & \cellcolor{gray!25}12:10 & \cellcolor{gray!25}3850  &\cellcolor{gray!25}CIR \\
 $18\_2003$ & 18 Jul & 16:56 & 21 Jul  & 07:51 & 3775  &?\\
 \cellcolor{gray!25}$19\_2003$ & \cellcolor{gray!25}27 Jul & \cellcolor{gray!25}09:00 & \cellcolor{gray!25}04 Aug  & \cellcolor{gray!25}10:46 & \cellcolor{gray!25}11628  &\cellcolor{gray!25}CIR \\
 $20\_2003$ & 07 Aug & 23:00 & 14 Aug  & 11:09 & 9369  &CIR \\
 \cellcolor{gray!25}$21\_2003$ & \cellcolor{gray!25}21 Aug & \cellcolor{gray!25}08:00 & \cellcolor{gray!25}25 Aug  & \cellcolor{gray!25}18:19 & \cellcolor{gray!25}6379  &\cellcolor{gray!25}CIR \\
 $22\_2003$ & 10 Sep & 08:02 & 12 Sep  & 17:39 & 3457 &CIR\\
 \cellcolor{gray!25}$23\_2003$ & \cellcolor{gray!25}18 Sep & \cellcolor{gray!25}00:00 & \cellcolor{gray!25}22 Sep  & \cellcolor{gray!25}20:02 & \cellcolor{gray!25}5522  &\cellcolor{gray!25}?\\
 $24\_2003$ & 24 Sep & 09:00 & 27 Sep  & 01:32 & 3872  &?\\
 \cellcolor{gray!25}$25\_2003$ & \cellcolor{gray!25}15 Oct & \cellcolor{gray!25}12:00 & \cellcolor{gray!25}22 Oct  & \cellcolor{gray!25}18:35 & \cellcolor{gray!25}10475  &\cellcolor{gray!25}CIR/ICME\\
 \cellcolor{gray!25}$26\_2003$ & \cellcolor{gray!25}01 Nov & \cellcolor{gray!25}07:00 & \cellcolor{gray!25}04 Nov  & \cellcolor{gray!25}06:16 & \cellcolor{gray!25}4278  &\cellcolor{gray!25}ICME\\
 $27\_2003$ & 09 Nov & 01:41 & 19 Nov  & 18:46 & 15425  &ICME/CIR\\
 $28\_2003$ & 07 Dec & 06:32 & 17 Dec  & 06:36 & 14404 &CIR \\
 \hline
 $01\_2004$ & 01 Jan & 04:00 & 06 Jan  & 11:11 & 3868  &CIR \\
 $02\_2004$ & 12 Jan & 00:01 & 15 Jan  & 05:31 & 3869  &ICME\\
 \cellcolor{gray!25}$03\_2004$ & \cellcolor{gray!25}15 Jan & \cellcolor{gray!25}14:06 & \cellcolor{gray!25}21 Jan  & \cellcolor{gray!25}23:58 & \cellcolor{gray!25}3057  &\cellcolor{gray!25}CIR \\
 \cellcolor{gray!25}$04\_2004$ & \cellcolor{gray!25}29 Jan & \cellcolor{gray!25}09:31 & \cellcolor{gray!25}31 Jan  & \cellcolor{gray!25}12:28 & \cellcolor{gray!25}5811  &\cellcolor{gray!25}ICME\\
 \cellcolor{gray!25}$05\_2004$ & \cellcolor{gray!25}12 Feb & \cellcolor{gray!25}00:01 & \cellcolor{gray!25}16 Feb  & \cellcolor{gray!25}00:52 & \cellcolor{gray!25}3272  &\cellcolor{gray!25}CIR \\
 $06\_2004$ & 28 Feb & 20:10 & 03 Mar  & 02:42 & 4287  &?\\
 \cellcolor{gray!25}$07\_2004$ & \cellcolor{gray!25}13 Mar & \cellcolor{gray!25}06:05 & \cellcolor{gray!25}16 Mar  & \cellcolor{gray!25}05:32 & \cellcolor{gray!25}4446  &\cellcolor{gray!25}CIR \\
  \cellcolor{gray!25}$08\_2004$ & \cellcolor{gray!25}06 Apr & \cellcolor{gray!25}08:00 & \cellcolor{gray!25}09 Apr  & \cellcolor{gray!25}10:06 & \cellcolor{gray!25}5445  &\cellcolor{gray!25}ICME+CIR \\
 $09\_2004$ & 02 May & 17:57 & 06 May  & 12:42 & 3764  &CIR \\
 $10\_2004$ & 28 May & 08:21 & 31 May  & 13:44 & 4644  &CIR \\
 $11\_2004$ & 05 Jun & 10:18 & 10 Jun  & 07:08 & 7010  &? \\
 $12\_2004$ & 20 Aug & 00:51 & 23 Aug  & 09:43 & 4852  &? \\
 $13\_2004$ & 06 Sep & 05:43 & 08 Sep  & 14:41 & 3418  &CIR \\
 $14\_2004$ & 13 Sep & 20:04 & 18 Sep  & 12:36 & 6753  &ICME\\
 $15\_2004$ & 03 Oct & 05:29 & 05 Oct  & 08:40 & 3071  &?\\
 $16\_2004$ & 12 Oct & 21:56 & 15 Oct  & 21:10 & 4275  &CIR \\
 $17\_2004$ & 27 Nov & 05:54 & 01 Dec  & 12:00 & 6126  &CIR \\
 $18\_2004$ & 06 Dec & 04:54 & 09 Dec  & 01:23 & 4110  &?\\
 \hline
 \cellcolor{gray!25}$01\_2005$ & \cellcolor{gray!25}01 Jan & \cellcolor{gray!25}19:07 & \cellcolor{gray!25}06 Jan  & \cellcolor{gray!25}00:00 & \cellcolor{gray!25}6053  &\cellcolor{gray!25}CIR?\\
 $02\_2005$ & 29 Jan & 06:13 & 01 Feb  & 00:56 & 4003  &CIR\\
 \cellcolor{gray!25}$03\_2005$ & \cellcolor{gray!25}07 Feb & \cellcolor{gray!25}22:00 & \cellcolor{gray!25}12 Feb  & \cellcolor{gray!25}05:24 & \cellcolor{gray!25}6204  &\cellcolor{gray!25}CIR?\\
 \cellcolor{gray!25}$04\_2005$ & \cellcolor{gray!25}07 Mar & \cellcolor{gray!25}02:00 & \cellcolor{gray!25}10 Mar  & \cellcolor{gray!25}18:02 & \cellcolor{gray!25}5282  &\cellcolor{gray!25}CIR\\
 \cellcolor{gray!25}$05\_2005$ & \cellcolor{gray!25}05 Apr & \cellcolor{gray!25}09:00 & \cellcolor{gray!25}07 Apr  & \cellcolor{gray!25} 16:26 & \cellcolor{gray!25}3326 & \cellcolor{gray!25} CIR\\
 \cellcolor{gray!25}$06\_2005$ & \cellcolor{gray!25}12 Apr & \cellcolor{gray!25}07:00 & \cellcolor{gray!25}16 Apr  & \cellcolor{gray!25}06:16 & \cellcolor{gray!25}5716  &\cellcolor{gray!25}CIR\\
 $07\_2005$ & 30 Apr & 03:01 & 04 May  & 06:31 & 5970  &CIR\\
 \cellcolor{gray!25}$08\_2005$ & \cellcolor{gray!25}11 May & \cellcolor{gray!25}06:55 & \cellcolor{gray!25}13 May  & \cellcolor{gray!25}17:57 & \cellcolor{gray!25}3542  &\cellcolor{gray!25}?\\
 $09\_2005$ & 28 Jul & 06:00 & 30 Jul  & 09:35 & 4535  &CIR\\
 $10\_2005$ & 05 Aug & 11:52 & 08 Aug  & 15:04 & 4512  &CIR\\
 $11\_2005$ & 02 Nov & 22:00 & 05 Nov  & 17:12 & 4032  &CIR\\
 $12\_2005$ & 12 Nov & 00:13 & 14 Nov  & 22:57 & 4244  &?\\
 $13\_2005$ & 27 Dec & 15:13 & 30 Dec  & 09:55 & 4002  &CIR\\
 \hline
 $01\_2006$ & 19 Feb & 15:40 & 22 Feb  & 20:55 & 4635  &CIR\\
 $02\_2006$ & 19 Mar & 01:24 & 22 Mar  & 09:58 & 4834  &CIR\\
 \cellcolor{gray!25}$03\_2006$ & \cellcolor{gray!25}15 Apr & \cellcolor{gray!25}00:00 & \cellcolor{gray!25}17 Apr  & \cellcolor{gray!25}01:48 & \cellcolor{gray!25}2988 &\cellcolor{gray!25}ICME\\
 $04\_2006$ & 06 Jun & 20:00 & 11 Jun  & 17:46 & 7066  &CIR\\
 $05\_2006$ & 14 Jun & 21:47 & 17 Jun  & 19:05 & 4158  &CIR\\
 \cellcolor{gray!25}$06\_2006$ & \cellcolor{gray!25}20 Aug & \cellcolor{gray!25}15:00 & \cellcolor{gray!25}22 Aug  & \cellcolor{gray!25}19:20 & \cellcolor{gray!25}3140  & \cellcolor{gray!25}ICME\\
 \cellcolor{gray!25}$07\_2006$ & \cellcolor{gray!25}14 Oct & \cellcolor{gray!25}02:00 & \cellcolor{gray!25}16 Oct  & \cellcolor{gray!25}23:06 & \cellcolor{gray!25}4146  & \cellcolor{gray!25}CIR\\
 \cellcolor{gray!25}$08\_2006$ & \cellcolor{gray!25}10 Nov & \cellcolor{gray!25}08:00 & \cellcolor{gray!25}12 Nov  & \cellcolor{gray!25}14:12 & \cellcolor{gray!25}3252  & \cellcolor{gray!25}CIR\\
 $09\_2006$ & 23 Nov & 16:00 & 27 Nov  & 20:25 & 6025  &CIR\\
 \cellcolor{gray!25}$10\_2006$ & \cellcolor{gray!25}06 Dec & \cellcolor{gray!25}13:00 & \cellcolor{gray!25}09 Dec  & \cellcolor{gray!25}04:27 & \cellcolor{gray!25}3807  &\cellcolor{gray!25}CIR/ICME\\
 $11\_2006$ & 19 Dec & 23:29 & 24 Dec  & 00:42 & 5833  &CIR\\
 \hline
 $01\_2007$ & 01 Jan & 23:03 & 04 Jan  & 18:26 & 4043  &CIR\\
 $02\_2007$ & 16 Jan & 09:39 & 20 Jan  & 01:58 & 5299  &CIR/ICME?\\
 $03\_2007$ & 29 Jan & 19:00 & 01 Feb  & 00:49 & 3229  &CIR\\
 $04\_2007$ & 27 Feb & 08:45 & 01 Mar  & 17:07 & 3382  &CIR\\
 \cellcolor{gray!25}$05\_2007$ & \cellcolor{gray!25}01 Apr & \cellcolor{gray!25}09:00 & \cellcolor{gray!25}03 Apr  & \cellcolor{gray!25}11:15 & \cellcolor{gray!25}3015  &\cellcolor{gray!25}CIR\\
 $06\_2007$ & 27 Apr & 09:48 & 30 Apr  & 16:59 & 4751  &CIR\\
 \cellcolor{gray!25}$07\_2007$ & \cellcolor{gray!25}23 May & \cellcolor{gray!25}15:00 & \cellcolor{gray!25}27 May  & \cellcolor{gray!25}20:18 & \cellcolor{gray!25}6078  &\cellcolor{gray!25}?\\
 $08\_2007$ & 10 Aug & 07:39 & 12 Aug  & 12:19 & 3160  &CIR\\
 $09\_2007$ & 01 Sep & 16:31 & 03 Sep  & 17:10 & 2919  &CIR\\
 $10\_2007$ & 05 Sep & 00:21 & 07 Sep  & 04:43 & 3142  &CIR\\
 $11\_2007$ & 24 Nov & 07:30 & 26 Nov  & 11:17 & 3107  &?\\
 \end{longtable}

\begin{table}
  \begin{center}
    \caption{Average and standard deviation of the AE index, and Time of occurrence of HILDCAA* ($\%$) between 1998-2007.}
    \label{tab:table2}
    \begin{tabular}{llll}
      \hline
      \textbf{Year} & \textbf{Average (nT)} & \textbf{Sd(nT)} & \textbf{HILDCAA* ($\%$)}  \\
      \hline
      1998  & 207.55 & 144.72 & 7.1\\
      1999  & 218.52 & 137.73 & 14.2\\
      \colorbox{brown!30}{2000}  & \colorbox{brown!30}{235.65} & \colorbox{brown!30}{150.70} & \colorbox{brown!30}{8.6}\\
      2001  & 202.84 & 141.98 & 7.7\\
      2002  & 216.96 & 146.85 & 12.2\\
     \colorbox{brown!50}{2003}  & \colorbox{brown!50}{328.26} & \colorbox{brown!50}{175.83} & \colorbox{brown!50}{42.2}\\
      2004  & 220.73 & 153.03 & 15.6\\
      2005  & 225.24 & 160.05 & 11.7\\
      2006  & 151.24 & 120.15 & 9.5\\
      2007  & 130.40 & 102.26 & 8.0\\
      \hline
    \end{tabular}
  \end{center}
\end{table}

\begin{table}
  \begin{center}
    \caption{Mean values of the parameters registered among all 135 HILDCAA* events.}
    \label{tab:table3}
    \begin{tabular}{ll}
      \hline
      \textbf{PARAMETERS} & \textbf{AVERAGE $\pm$ $\sigma$ } 	 \\
      \hline
      AE (nT)              & $348.1\pm67.1$ \\
     AU (nT)               & $123.9\pm34.2$ \\
     AL (nT)               & $-224.2\pm47.3$ \\
     Dst (nT)              & $-22.1\pm9.1$ \\
     Density ($cm^{-3}$)   & $4.3\pm1.3$ \\
     Temperature (K)       & $170976.3\pm58049.9$ \\
     $|V|$ (km/s)            & $547.1\pm83.9$ \\
      Bz (nT)              & $-0.70\pm0.88$ \\
     $|B|$ (nT)              & $6.6\pm1.2$ \\
     Pressure (nPa)        & $2.4\pm0.6$ \\
     Beta                  & $0.66\pm0.27$ \\
      \hline
    \end{tabular}
  \end{center}
\end{table}

\begin{table}
\begin{center}
    \caption{Solar wind parameters related to HILDCAA* events compared with different levels of geomagnetic activity.}
    \label{tab:table4}
\begin{tabular*}{1.0\textwidth}{@{\extracolsep{\fill}}p{5cm}  @{}l@{}  @{}l@{}  @{}l@{}}
  \hline
  \multicolumn{4}{c}{\textbf{Higher values during HILDCAA*}} \\[0.2ex] 
\cline{1-4}
    \multicolumn{1}{ r }{}& \textbf{\tiny{Background}} & \textbf{\footnotesize{HILDCAA*}} & \textbf{\tiny{High Activity (geomagnetic Storm)}}\\
  \hline
Tp(K)      & $\approx 1.1\times10^5$  &$>1.1\times 10^5$     &$<1.1\times 10^5$\\
V (km/s)   & $<400$                   &$>450$                &$\approx 450$\\
Beta       &                          &$\approx 0.7$           &$\ll 1$\\
\cline{1-4}
\multicolumn{4}{c}{\textbf{Intermediates}} \\[0.2ex] 
\cline{1-4}
Bs (nT)    & $0$             &$-2\le Bs\le 0$  &$<-3$\\
Bo (nT)   & $\approx 5$     &$5< Bo< 9$    & $>9$\\
\cline{1-4}
\multicolumn{4}{c}{\textbf{Smaller}} \\[0.2ex] 
\cline{1-4}
Density ($cm^{-3}$)& $\approx 6.1$&$< 6$&$\gg 6$\\
  \hline
\end{tabular*}
\end{center}
\end{table}

\newpage

\end{document}